\documentclass[a4paper,11pt]{article}
\usepackage{pos}
\usepackage{amsmath}
\usepackage[T1]{fontenc}
\usepackage{autobreak}
\usepackage[english]{babel}

\newcommand{\myspace}{\,}

\newcommand{\msbar}{\overline{\mathrm{MS}}}

\title{Single- and double-unresolved limits of polarized tree-level matrix elements}

\author{Thomas Gehrmann}
\author*{Markus Löchner}

\affiliation{Physik-Institut, Universität Zürich,\\
  Winterthurerstrasse 190, 8057 Zürich, Switzerland}

\emailAdd{thomas.gehrmann@uzh.ch}
\emailAdd{markus.loechner@physik.uzh.ch}

\abstract{The calculation of exclusive cross sections at next-to-next-to-leading order (NNLO) in QCD requires an analytic understanding of the infrared singular structure with up to two unresolved partons. This has so far only been achieved for unpolarized matrix elements. We derive the full set of splitting amplitudes arising in longitudinally polarized tree-level QCD matrix elements at NNLO in the Larin $\gamma_5$ scheme. They are extracted from DIS-like processes, and are verified in matrix elements of higher multiplicity. Our results will enable the calculation of NNLO corrections to longitudinal spin asymmetries in polarized collider processes.}

\FullConference{31st International Workshop on Deep Inelastic Scattering (DIS2024)\\
 8–12 April 2024\\
Grenoble, France\\}

\begin{document}
\maketitle

\section{Introduction}
The proton structure 
in terms of parton distributions (PDFs)  is known to the level of a few percent \cite{Hou:2019efy,Bailey:2020ooq,NNPDF:2021njg} for the case of unpolarized PDFs. These describe the 
momentum distribution among its partonic constitutents: quarks and gluons. If the proton is spin-polairzed, one can also define 
polarized PDFs, which describe  difference of momentum distributions of partons with spin aligned and 
anti-aligned with the proton spin. With much less data available on high-energy collisions involving 
polarized protons~\cite{Aidala:2012mv}, 
the knowledge on polarized PDFs is considerable less precise on their unpolarized counterparts, see e.g.~\cite{Borsa:2024mss}.  
This  experimental bottleneck promises to finally be overcome with the upcoming Electron-Ion Collider (EIC) at BNL \cite{Aschenauer:2014cki,Aschenauer:2019kzf}, which will measure a multitude of polarized observables  to high 
statistical precision. The upcoming EIC dataset calls for substantial improvements to theory predictions for 
polarized observables, to avoid that physics studies with EIC data will become limited by theory uncertainties.

The theory description of spin effects introduces anti-symmetric structures like the $\gamma_5$ matrix and Levi-Civita tensors into the calculation, which are inconsistent objects in dimensional regularization in $d=4-2\varepsilon \neq 4$ dimensions. Several prescriptions were put forward to  fix this inconsistency \cite{tHooft:1972tcz,Breitenlohner:1977hr,Korner:1991sx,Larin:1993tq},
 but often they severely obstruct the calculation of cross sections. From a practical point of view, the Larin scheme \cite{Larin:1993tq} appears to be best-suited for higher order perturbative calculations. 
 It merely increases the algebraic complexity of the calculations, but in turn breaks the axial current Ward identities, which 
 need to be restored by appropriate renormalization.

Inclusive results for polarized deep-inelastic scattering (DIS) at NNLO have been available for decades \cite{Zijlstra:1993sh}, while less inclusive observables at NNLO for fragmentation~\cite{Bonino:2024wgg,Goyal:2024tmo} and single jet production~\cite{Borsa:2020ulb,Borsa:2020yxh} are only emerging recently. The former could still be integrated analytically, whereas the latter require the use of numerical integration techniques, based on the 
Projection-to-Born method~\cite{Cacciari:2015jma} combined with dipole subtraction~\cite{Catani:1996vz}. For even more involved jet observables at NNLO, a general NNLO subtraction scheme for polarized processes has yet to be formulated. For doing that, full knowledge of the IR structure in tree-level double-unresolved and one-loop corrected single-unresolved limits of generic polarized matrix elements is required.

Subtraction schemes rely on the factorization of the real-radiation matrix element in the infrared singular limits into a universal object called the splitting amplitude for collinear limits, or eikonal factor in the soft limit. It is a priori unclear if such a factorization also exists for polarized matrix elements. From the analysis of DIS-like tree-level real radiation matrix elements with up to five partons, we present convincing evidence for factorization: we are able to extract the universal splitting amplitudes and formulate systematic rules for the factorization of polarized tree-level matrix elements in a consistent $\gamma_5$ scheme at NNLO.

In section~\ref{sec:method} we outline the general method for obtaining infrared limits from matrix elements. In sections~\ref{sec:single_unresolved} and~\ref{sec:double_unresolved} we describe the singular limits present in tree-level matrix elements up to NNLO, and comment in detail on how factorization of the matrix element is obtained.

\section{Method}
\label{sec:method}
The unpolarized quark-initiated triple-collinear splitting amplitudes were first extracted from the tree-level matrix element for $\gamma^*\to 4~\mathrm{partons}$, and checked numerically against the factorization of  matrix element $\gamma^*\to 5~\mathrm{partons}$ with one additional real emission.  The latter also granted access to gluon-initiated triple-collinear splitting amplitudes \cite{Campbell:1997hg}. 

A generic proof of factorization for $n$-parton splitting amplitudes, which also contained a straightforward method for the calculation of splitting amplitudes, was provided subsequently~\cite{Catani:1999ss}. It also captured the azimuthal terms, and provided a straightforward method for the calculation of general $m$-parton splitting amplitudes. After azimuthal averaging~\cite{Catani:1998nv} the triple-collinear splitting amplitudes of \cite{Campbell:1997hg} are recovered.

Although such a constructive proof would be desirable, this approach fails in case of polarized splitting amplitudes due to the intricacy of correctly parametrizing off-shell internal polarized partonic currents. Instead, we here resort to the extraction from the unintegrated tree-level contributions to the (crossed) $g_1$ coefficient function with $\gamma^* \to n~\mathrm{partons}$, cf.~\cite{Zijlstra:1993sh}, as well as the graviton-initiated $g_1$ contributions with $G^*\to  n~\mathrm{partons}$~\cite{Lam:1981tg,Stirling:2011tf,Moch:2014sna}, with $n=3,4$ for single- and double-unresolved limits respectively. Only massless partons are considered. For a consistent treatment of $\gamma_5$ we use the Larin scheme \cite{Larin:1993tq}.

The inclusion of graviton matrix elements allows also for a check for polarized gluon-initiated splitting amplitudes at one order higher, which would otherwise only be available with an additional radiation in photon matrix elements. We check the limits obtained this way up to $n=5$ for the photon and $n=4$ for the graviton decay, obtaining analytical factorization in each single- and double-unresolved limit.

Analytical factorization is finally achieved by recasting the analytical expression of the limit into a unique form, obtained by either factorizing the ensuing rational functions, or by partial fraction decomposition with help of the Le\u{i}nartas decomposition \cite{Leinartas:1978,Heller:2021qkz} in \textsc{FORM} 4 \cite{Kuipers:2012rf}.

For $n=3,4$ we also determine the limits for an initial-state polarized parton with momentum parametrizations different from the ones presented here, retaining the results from the all-final case multiplied by a universal Jacobian factor and trivial factors depending on the crossed particle.

\section{Single-unresolved limits}
\label{sec:single_unresolved}
Single-unresolved limits are present in any tree-level DIS-like matrix element with parton multiplicity $n\geq 3$. In this section we describe the universal structures arising in the single-unresolved limit of tree-level matrix elements. They are extracted from the case $n=3$, where apart from the collinear momentum fraction $z$ there is only one kinematic scale $q^2=(p_1+p_2+p_3)^2$, making factorization trivial. The singular limits are then reproduced for multiplicities $n=4$ and $n=5$, where the factorization is non-trivial, providing a strong check. Since the tree-level matrix elements are singular objects in the unresolved regions of the phase space, subleading terms in the dimensional regularization parameter $\varepsilon$ in $d=4-2\varepsilon$ dimensions must be retained.


Let us consider two massless momenta $p_i$, $p_j$, and $\vartheta = \angle(\vec p_i,\vec p_j) $. A propagator $1/s_{ij}$ is then written as
\begin{align}
\frac{1}{s_{ij}} = \frac{1}{E_i E_j (1-\cos \vartheta)}.
\end{align}
In this form it is apparent that the propagator becomes divergent when either $E_i,E_j\to 0$ (Sec.~\ref{sec:single_soft}), or in the limits $\vartheta\to 0$, $\vartheta\to\pi$ (Sec.~\ref{sec:single_collinear}). Due to gauge invariance, only single poles in a single invariant may develop in the final result, whereas higher poles are always spurious.

\subsection{Single-soft limits}
\label{sec:single_soft}
The single-soft limit is kinematically non-trivial for gluons only. It is obtained by rescaling the gluon momentum by $p_j \to \lambda p_j$ and taking the limit $\lambda\to 0$. An $m+1$ parton matrix element 
with gluon $j$ unpolarized then factorizes down to an $m$-parton matrix element and an eikonal factor,
\begin{align}
\Big|(\Delta)\mathcal{M}_{m+1}^{(0)}&(p_1,\dots, p_i, \lambda p_j, p_k, \dots,p_{m+1})\Big|^2 \nonumber \\ &\overset{\lambda \to 0}{\longrightarrow} 4\pi\alpha_s \frac{1}{\lambda^2} \frac{2s_{ik}}{s_{ij}s_{jk}} \left|(\Delta)\mathcal{M}_m^{(0)}(p_1,\dots, p_i, p_k, \dots , p_{m+1})\right|^2 + \mathcal{O}\big(\lambda^0\big), \label{eq:single_collinear_fact}
\end{align}
where $s_{i_1\dots i_k} = (p_{i_1}+ \,\cdots \,+ p_{i_k})^2$. 
If gluon $j$ is polarized, the matrix element remains non-singular in the soft limit. 
This behaviour relates to the non-singular limit $z\to 0$ in $\Delta P_{gq}(z)$ and $\Delta P_{gg}(z)$  in Eq.~\eqref{eq:single_collinear_splitting} below, which is singular in the unpolarized splitting amplitude.

\subsection{Single-collinear limits}
\label{sec:single_collinear}
The single-collinear limit between partons $i$, $j$ is understood in a light cone parametrization of their momenta,
\begin{align}
p_i^\mu =& z p^\mu + k_{T}^\mu - \frac{k_{T}^2}{2 z}\frac{n^\mu}{n\cdot p},  &
p_j^\mu =& (1-z) p^\mu - k_{T}^\mu - \frac{k_{T}^2}{2 (1-z)}\frac{n^\mu}{n\cdot p}, \label{eq:light_cone_param_single}
\end{align}
in the limit where the transverse momentum $k_T$ tends to zero, and where 
\begin{align}
p^2 &= 0= n^2, & k_{T}^2 &< 0, &
 p\cdot k_{T} &= 0 = n\cdot k_T, &
 n\cdot p \neq 0. \label{eq:light_cone_momenta_rels}
\end{align}

After insertion of the light cone parametrization inside a matrix element, a dependence on the transverse momentum $k_{T}$ may remain, either in terms of angular correlations, e.g. $k_{T}^\mu k_T^{\nu}$, or contracted with scalar products. However, as $k_T$ is entirely unresolved in the limit considered, it needs to be integrated over. Therefore we can make a tensor basis ansatz, whose solution is given by
\begin{align}
&\int \mathrm{d}\varphi\, k_T^{\mu}(\varphi) = 0, \nonumber \\
&\int \mathrm{d}\varphi\, k_T^{\mu}(\varphi) k_T^{\nu}(\varphi) = \frac{k_{T}^2}{d-2} \left( -g^{\mu\nu} + \frac{p^{\mu} n^{\nu} + p^\nu n^\mu}{p\cdot n}\right). \label{eq:transverse_momentum_projection_scol}
\end{align}
Notice that only symmetric tensor structures are allowed here.
The matrix element then factorizes in the collinear limit between $i$ and $j$ into
\begin{align}
\left|(\Delta)\mathcal{M}_{m+1}^{(0)}(p_1,\dots,p_i,p_j,\dots p_{m+1})\right|^2 \overset{i\parallel j}{\longrightarrow} \, &  4\pi \alpha_s \frac{ \mathcal{P}_{ij \leftarrow a}(z)}{s_{ij}} \, |(\Delta)\mathcal{M}_m^{(0)}(p_1,\dots, p, \dots, p_{m+1})|^2 \nonumber \\
& + \mathcal{O}(k_T^0)
\end{align}
where $ \mathcal{P}_{ij \leftarrow a}(z) $ is to be read as $\mathbf{P}_{ij \leftarrow a}(z)$ if $i$ and $j$ are unpolarized particles, and as $\mathbf{P}_{\Delta ij \leftarrow a}(z)$ if $i$ is  polarized. In the latter case, the merged particle $a$ with momentum $p$ in the reduced matrix element $\mathcal{M}_m^{0}$ is also polarized. Here, $a$ denotes the particle initiating the splitting.

The full set of tree-level single-collinear splitting amplitudes in polarized processes is given by
\begin{align}
\mathbf{P}_{(\Delta) q\, g \leftarrow q} &= C_F P_{(\Delta) q\, g \leftarrow q}, &
\mathbf{P}_{(\Delta) g \,q \leftarrow q} &= C_F P_{(\Delta) g \,q \leftarrow q} , \nonumber \\
\mathbf{P}_{(\Delta) q \, q \leftarrow g} &= T_R P_{(\Delta) q \, q \leftarrow g}, &
\mathbf{P}_{(\Delta) g \, g \leftarrow g} &= C_A P_{(\Delta) g \, g \leftarrow g},
\end{align}
with the color-stripped splitting amplitudes defined as
\begin{align}
P_{qg \leftarrow q}(z) &= \frac{1+z^2}{1-z}-(1-z)\varepsilon, & P_{\Delta q \, g \leftarrow q}(z) &= 
\frac{1+z^2}{1-z}+(1-z)\frac{\varepsilon(3+\varepsilon)}{1-\varepsilon}, \nonumber \\
P_{gq \leftarrow q}(z) &= \frac{1+(1-z)^2}{z} - z \varepsilon , & P_{\Delta g \,q \leftarrow q}(z) &= \frac{2-z(1+\varepsilon)}{1-\varepsilon} , \nonumber \\
P_{qq \leftarrow g}(z) &=  1-2 \frac{z(1-z)}{1-\varepsilon}, & P_{\Delta q \, q \leftarrow g}(z) &=  1 - 2 \frac{1-z}{1-\varepsilon} , \nonumber \\
P_{gg \leftarrow g}(z) &=  2 z(1-z) + \frac{2}{1-z}+\frac{2}{z}-4 , &  P_{\Delta g \, g \leftarrow g}(z) &= \frac{2}{1-z} -2 + 4 \frac{1- z}{1-\varepsilon}. \label{eq:single_collinear_splitting}
\end{align}
Here, quarks may also be exchanged by antiquarks. We notice the genuinely different structure if a polarized particle is involved in the splitting. 

The tree-level single-collinear splitting amplitudes $\mathcal{P}_{ij}$ are given by the unregulated real radiation part of the corresponding NLO Altarelli-Parisi evolution kernel $\Delta P_{ia}$~\cite{Altarelli:1977zs} and an evanescent $\varepsilon$-dependent term. The latter depends on the scheme choice for $\gamma_5$. This becomes apparent for the non-singlet splitting \(\displaystyle P_{\Delta q\, g \leftarrow q }(z) = \frac{1+z^2}{1-z} \, +  \, 3(1-z)\varepsilon \, + \, \mathcal{O}(\varepsilon^2) \) by considering the finite scheme transformation required to get from the Larin scheme to $\msbar$~\cite{Matiounine:1998re}, fixing the axial Ward identity and restoring helicity conservation along the polarized quark line, after which $P_{\Delta q \, g \leftarrow q}$ takes the form of $P_{qg \leftarrow q}$.

\section{Double-unresolved limits}
\label{sec:double_unresolved}
Due to the more involved singular structure in the case of the double-unresolved limits appearing in the double-real radiation part at NNLO, we resort to color-ordered matrix elements as in \cite{Campbell:1997hg}, containing singular limits only between color-adjacent particles~\cite{Berends:1987cv,Mangano:1987xk,Mangano:1988kk,Bern:1990ux}. Note that the insertion of a colorless particle into the amplitude does not modify the color structure, independent of where it couples. In the context of color-ordered amplitudes the notion of abelian gluons is introduced, referring to unordered gluons, which 
can be  color-adjacent 
only to quarks but not to gluons. 

\subsection{Double-soft limits}
We parametrize the two momenta $p_i$, $p_j$ as in the single-soft case by rescaling with a parameter~$\lambda$. The double-soft limit is then realized in the limit $\lambda\to 0$.

In the case of color-unconnected particles, the matrix element factorizes into two separate eikonal factors.  For color-connected particles, we distinguish three cases: two soft non-abelian gluons, two soft abelian gluons, and two soft quarks. We recover the limits quoted in~\cite{Braun-White:2023sgd} if neither soft particle is polarized.

As before, if a polarized particle becomes soft the matrix element does not develop a singularity. 

\subsection{Double-collinear limits}
If the parton momentum $p_i$ becomes collinear to $p_j$, and $p_k$ becomes collinear to $p_l$, the matrix element factorizes into two separate single-collinear splitting amplitudes.

\subsection{Triple-collinear limits}
In the triple-collinear case, where a cluster of three particles becomes collinear, we employ the light-cone parametrization
\begin{align}
p_i^\mu=z_i p^\mu + k_{T,i}^\mu - \frac{k_{T,i}^2}{2 z_i}\frac{n^\mu}{n\cdot p}. \label{eq:tcol_sudakov}
\end{align}
As discussed in \cite{Catani:1998nv}, the relations 
\begin{align}
z_1 + z_2 + z_3 &= 1, &
 \sum_{i=1}^3 k_{T,i} = 0
\end{align}
hold. With a similar power counting as before, this time discarding terms of $\mathcal{O}(k_T^{-3})$ and less singular, we find a residual transverse momentum dependence of the form $\{k_{T,i}^\mu, k_{T,i}^\mu k_{T,i}^\nu, k_{T,i}^\mu k_{T,j}^\nu\}$. The first and second case are as in Eq.~\eqref{eq:transverse_momentum_projection_scol}. In the third case, however, the tensor ansatz changes to 
\begin{align}
\int\mathrm{d}\phi_{ij} \, k_{T,i}^\mu k_{T,j}^\nu = A g^{\mu \nu} + B p^\mu p^\nu + C n^\nu n^\mu + D p^\mu n^\nu + E n^\mu p^\nu + F \varepsilon^{\mu\nu\rho\sigma}\frac{p_\rho n_\sigma}{p\cdot n},
\end{align} 
where $\phi_{ij} = \angle (k_{T,i}, k_{T,j})$. It turns out that the symmetric part
$\int\mathrm{d}\phi_{ij} \, k_{T,i}^{(\mu} k_{T,j}^{\nu)}$ takes the overall form of Eq.~\eqref{eq:transverse_momentum_projection_scol}, 
\begin{align}
\int \mathrm{d}\varphi\, k_{T,i}^{(\mu}(\varphi) k_{T,j}^{\nu)}(\varphi) = \frac{\sqrt{k_{T,i}^2 k_{T,j}^2}}{d-2} \left( -g^{\mu\nu} + \frac{p^{\mu} n^{\nu} + p^\nu n^\mu}{p\cdot n}\right),
\end{align}
and gives rise to an angular dependence proportional to $\cos(2\phi_{ij})$ in 4-dimensions. The antisymmetric counterpart is found to have a $\sin(2\phi_{ij})$ dependence of the opening angle between the collinear partons, and its coefficient $F$ is regularization scheme dependent. Note that the two structures are mutually orthogonal to each other. However, while it is in principle allowed, we find that the antisymmetric part is not realized in the tree-level matrix elements required up to di-jet production at NNLO.

After the angular averaging, our matrix elements factorize according to
\begin{align}
\vert (\Delta) &\mathcal{M}_{m+2} (p_1,\dots, p_i, p_j, p_k,\dots,p_{m+2}) \vert^2 \nonumber \\ & \overset{i\parallel j}{\longrightarrow} (4\pi \alpha_s)^2 \frac{1}{s_{ijk}^2} \mathcal{P}_{ijk}(z_i,z_j,z_k,s_{ij},s_{ik},s_{jk}) \times
\vert \mathcal{M}_{m} (p_1,\dots, p,\dots, p_{m+2} ) \vert^2 + \mathcal{O}\Bigg( \frac{1}{k_T^2}\Bigg),
\end{align}
where $\mathcal{P}_{ijk \leftarrow a}$ denotes the color-ordered triple-collinear splitting amplitude. In the unpolarized case there are 7 independent triple collinear splitting amplitudes \cite{Campbell:1997hg,Catani:1998nv},
$$
  P_{ggg  \leftarrow g},\,
  P_{qgg  \leftarrow q},\,
  P_{q\gamma \gamma  \leftarrow q},\,
  P_{gq\bar{q}  \leftarrow g},\,
  P_{\gamma q \bar{q} \leftarrow g},\,
  P_{q\bar{q}^\prime q^\prime  \leftarrow q},\,
  P_{q\bar{q}q  \leftarrow q}.
  $$
The polarized case is found to contain 16 additional independent splitting amplitudes, where one of the particles is polarized, indicated by a subscript~${}_\Delta$:
\begin{eqnarray*}
&   P_{\Delta q\myspace g \myspace g  \leftarrow q},\,
  P_{\Delta q\myspace \gamma \myspace\gamma  \leftarrow q},\,
  P_{\Delta g\myspace g\myspace q  \leftarrow q},\,
  P_{g\myspace \Delta g \myspace q  \leftarrow q},\,
  P_{\Delta \gamma\myspace \gamma \myspace q \leftarrow q},\,
  P_{\Delta q\myspace q^\prime\myspace \bar{q}^\prime \leftarrow q},\,
  P_{\Delta q^\prime \myspace \bar{q}^\prime q \leftarrow q},\,
  P_{\Delta q \myspace \bar{q} \myspace q \leftarrow q},\, \\
&    P_{\Delta \bar{q} \myspace q \myspace q \leftarrow q},\,
  P_{\Delta g \myspace g \myspace g \leftarrow g},\,
  P_{g \myspace \Delta g \myspace g \leftarrow g},\,
  P_{g \myspace \Delta q \myspace \bar{q} \leftarrow g},\,
  P_{\Delta q \myspace \bar{q} \myspace g \leftarrow g},\,
  P_{\Delta q \myspace \bar{q} \myspace \gamma \leftarrow g},\,
  P_{\Delta g \myspace q \myspace \bar{q} \leftarrow g},\,
  P_{\Delta \gamma \myspace q \myspace \bar{q} \leftarrow g}.
\end{eqnarray*}
For sake of abbreviation, they are denoted by $\mathcal{P}_{123 \to i}(z_1,z_2,z_3,s_{12},s_{13},s_{23})=P_{ijk \leftarrow a}$, where $i,j,k$ refer to the particle types of the clustered particles $1,2,3$, and $a$ is the type of the splitting particle. In this notation $\gamma$ stands for an abelian gluon.
By convention, triple-collinear splitting amplitudes are given in the literature with all particles in the final state, cf. Eq.~\eqref{eq:tcol_sudakov}.

Notice the different ordering of the clustered particles, arising from the breaking of symmetry between a $gg$ or a $q\bar{q}$ pair within the cluster, where either particle of the pair is polarized: in $P_{\Delta g\myspace g\myspace q  \leftarrow q}$ the unpolarized gluon is color-adjacent to the quark, whereas in $P_{g\myspace \Delta g \myspace q  \leftarrow q}$ the polarized gluon is. In $P_{g \myspace \Delta g \myspace g \leftarrow g}$, the polarized gluon is sandwiched between the two unpolarized gluons, which are symmetric under exchange, but in $P_{\Delta g \myspace g \myspace g \leftarrow g}$ it is adjacent to only one particular, breaking the symmetry. In $P_{g \myspace \Delta q \myspace \bar{q} \leftarrow g}$, the gluon is adjacent to the polarized quark and color-unconnected to the antiquark, and oppositely in $P_{\Delta q \myspace \bar{q} \myspace g \leftarrow g}$.
 
While triple-collinear splitting amplitudes are genuinely double-unresolved structures, they also contain iterated single-unresolved structures~\cite{Braun-White:2022rtg}. To make this manifest, let us consider the triple-collinear splitting of a polarized quark into a polarized quark and two gluons,
{\allowdisplaybreaks
\begin{align}
 \begin{autobreak}
\input{tcol.out}
 \end{autobreak}
 \label{eq:PDqgg}
\end{align}
}
setting $\mathrm{Tr}(ijkl)=z_i s_{jk} - z_j s_{ik} + z_k s_{ij}$, and 
\begin{align}
W_{ij} = (x_i s_{jk} - x_j s_{ik})^2 - \frac{2}{1-\varepsilon} \frac{x_i x_j x_k}{1-x_k}s_{ij}s_{ijk}.
\end{align}

The triple-collinear splitting amplitude is given in the basis proposed in~\cite{Braun-White:2022rtg}, serving to separate terms that are singular in the single-collinear limit from the remainder, which is regular in the single-collinear limit. 
The first two terms display an iterated application of single-collinear splittings, also making manifest the divergent structure of polarized matrix elements as follows.

The first term describes the part which is also divergent in the sole $s_{23}\to 0$ limit, encompassing the splitting of two gluons from a gluon after the splitting of a collinear gluon off a polarized quark. Since both gluons are unpolarized, also the corresponding gluonic splitting is unpolarized, and the polarization is fully contained in the polarized $m+1$ parton matrix element, which in turn factorizes down to a polarized splitting amplitude and a polarized $m$ parton matrix element with polarized clustered momentum $p$.

The second term represents the radiation of the unpolarized gluon $p_2$ off a polarized quark with clustered momentum $p_{12}$, and then a subsequent radiation of the unpolarized gluon $p_3$ off the polarized clustered quark $(12)$, where the reduced matrix element in every step is polarized with a polarized leg $p_{12}$ or $p$. Note that there is a consecutive limit $s_{13}\to 0$ and $s_{123}\to 0$ due to color ordering. The remainder contains only genuine triple-collinear divergences and double-soft divergences.

Based on our findings, we formulate the rules for the factorization of a collinear limit in polarized matrix elements:
\begin{enumerate}
\item Check if a polarized particle is contained in the collinear cluster. If so, the splitting amplitude is polarized.
\item The polarized reduced matrix element has a particle with clustered momentum $p$ inserted in place of the particle initiating the cluster, which is polarized if and only if the splitting was polarized.
\end{enumerate}

Color-summed (unordered) matrix elements are recovered from the color-ordered matrix elements by summing over the permutation of particles fixed by the color ordering. For unordered matrix elements, the following nine unordered triple-collinear splitting amplitudes are in place \cite{Catani:1998nv}, which can be obtained from the color-ordered ones by
{\allowdisplaybreaks
\begin{alignat}{3}
& \mathbf{P}_{\Delta q \, g_1 \, g_2 \leftarrow q} &&=
  \frac{C_F C_A}{2} \big[ {P}_{\Delta q \, g_1 \, g_2 \leftarrow q} + {P}_{\Delta q \, g_2 \, g_1 \leftarrow q} - P_{\Delta q \, \gamma_1 \, \gamma_2 \leftarrow q} \big] 
&&  + C_F^2 P_{\Delta q \, \gamma_1 \, \gamma _2 \leftarrow q} \nonumber \\
&\mathbf{P}_{\Delta g \, g \, q \leftarrow q} &&=
  \frac{C_F C_A}{2}\big[ P_{\Delta g \, g \, q \leftarrow q} + P_{ g \, \Delta g \, q \leftarrow q} - P_{\Delta \gamma \, \gamma \, q \leftarrow q} \big]
&&  + C_F^2 P_{\Delta \gamma \, \gamma \, q \leftarrow q}  \nonumber \\
& \mathbf{P}_{\Delta q\, q^\prime\, \bar{q}^\prime \leftarrow q} &&= T_R C_F P_{\Delta q\, q^\prime \, \bar{q}^\prime \leftarrow q} \nonumber\\
& \mathbf{P}_{\Delta q^\prime \, \bar{q}^\prime \, q \leftarrow q} &&= T_R C_F P_{\Delta q^\prime \, \bar{q}^\prime \, q \leftarrow q} \nonumber \\
& \mathbf{P}_{\Delta q \, \bar{q} \, q \leftarrow q} &&=  C_F \left( C_F -\frac{C_A}{2}\right) {P}_{\Delta q \, \bar{q} \, q \leftarrow q} \nonumber \\
& \mathbf{P}_{\Delta \bar{q} \, q \, q \leftarrow q} &&=  C_F \left( C_F -\frac{C_A}{2}\right) {P}_{\Delta \bar{q} \, q \, q \leftarrow q} \nonumber \\
& \mathbf{P}_{\Delta g \, q \, \bar{q} \leftarrow g} &&= 
  \frac{C_A T_R}{2} \big[ P_{\Delta g \, q \, \bar{q} \leftarrow g} + (2 \leftrightarrow 3) - P_{\Delta \gamma \, q \, \bar{q} \leftarrow g} \big] 
&&  +  C_F T_R {P}_{\Delta \gamma \, q \, \bar{q} \leftarrow g} \nonumber \\
& \mathbf{P}_{\Delta q \, \bar{q} \,  \gamma \leftarrow g } &&= 
  \frac{C_A T_R}{2}  \big[ {P}_{\Delta q \, \bar{q} \, g \leftarrow g} + ( 1 \leftrightarrow 2 ) - {P}_{\Delta q \, \bar{q} \, \gamma \leftarrow g} \big]
&&  + C_F T_R P_{\Delta q \, \bar{q} \,  \gamma \leftarrow g }  \nonumber \\
& \mathbf{P}_{\Delta g_1 \, g_2 \, g_3 \leftarrow g } &&=  \frac{C_A^2}{4} \big[  \phantom{+} P_{\Delta g_1 \, g_2 \, g_3 \leftarrow g} + ( 1\leftrightarrow 3) \nonumber \\[-1mm]
&&&  \hspace{1cm} + P_{\Delta g_1 \, g_3 \, g_2 \leftarrow g} + ( 1\leftrightarrow 2)  \nonumber \\
&&&  \hspace{1cm}  + P_{g_2 \, \Delta g_1 \, g_3 \leftarrow g} + ( 2 \leftrightarrow 3 ) \big] \,. 
\end{alignat}
}

\section{Conclusion}
We presented the unresolved limits of polarized tree-level matrix elements up to NNLO. We showed analytically that subtraction is feasible in the Larin scheme for the tree-level real-radiation contributions to DIS jet observables up to di-jet at NNLO. The full set of unresolved limits up to NNLO for any polarized matrix element will be presented in
a future publication~\cite{PolarizedUnresolvedLimitsNNLO}. 

\acknowledgments
We would like to thank Oscar Braun-White and Nigel Glover for discussions. This work has received funding from the Swiss National Science Foundation (SNF) under contract 200020-204200.

\bibliographystyle{JHEP}
\bibliography{bibliography}

\end{document}